\documentclass[12pt]{article}
\usepackage{amsmath}
\usepackage{amsfonts}
\usepackage{amssymb}
\usepackage[english]{babel}
\usepackage{times}

\usepackage[T1]{fontenc}
\usepackage{epsfig}
\usepackage{amsmath}
\usepackage{amssymb}

\begin{document}
	
	\title{Axial momentum and  quantization of the Majorana field }

	\author{H. Arod\'z  \\ 
		{\small
			Institute of Theoretical Physics, Jagiellonian University, Cracow, Poland}\footnote{henryk.arodz@uj.edu.pl}}
	\date{$\;$}
	
	\maketitle

	\begin{abstract}
New approach to quantization of the relativistic Majorana field is presented. It is based on expansion of the field into eigenfunctions of the axial momentum -- a novel observable introduced recently. Relativistic invariance is used as the main guiding principle instead of canonical formalism.  Hidden structure of the quantized Majorana field in the form of real Clifford algebra of Hermitian fermionic operators is unveiled. All generators of the Poincar\'e transformations 
 are found as solutions of certain operator equations, without invoking the principle of correspondence with classical conserved quantities.   Also
operators of parity  $\hat{\mbox{P}}$ and time reversal $\hat{\mbox{T}}$ are constructed.  

  \vspace*{4cm}

\end{abstract}

	\pagebreak

\section{Introduction}

The  Majorana field is a very interesting object of theoretical studies for several  reasons. First, it is the most fundamental  fermionic field. The  more popular Dirac field is in fact composed  of two Majorana fields \cite{Maj}. Furthermore, it is still not excluded  by experiments that neutrinos are quanta  of the Majorana field, see, e.g.,  \cite{part1}. In various attempts to go `beyond the Standard Model' the Majorana neutrinos play an important role. Recently,  Majorana fermions have been discussed widely also in condensed matter physics, see, e.g.,  \cite{cond1}, \cite{Aguado}. 

Quantized Majorana field was introduced already in the pioneering work \cite{Maj}, straightaway in a remarkably modern form, after a noteworthy discussion of pseudoclassical Majorana field in terms of  bispinors with anticommuting components, including the Hamiltonian formulation. 
As regards recent theoretical works involving the quantized  Majorana field,  we would like to mention  works on description of neutrino flavor oscillations, see, e.g.,  \cite{Blasone} and references therein, and works on the discrete symmetries $\mbox{P}$ and $\mbox{PC}$, see, e.g., \cite{Fujikawa1} and references therein.

Our paper is devoted to quantization of a classical Majorana field. Let us stress that this classical field should not be confused with the  more popular  pseudoclassical  Majorana field  which has anticommuting components and is considered already in \cite{Maj}.  In our paper the pseudoclassical field is not discussed. 
The classical Majorana field has commuting components. It
can be considered in three equivalent forms: as a four component complex Dirac bispinor subject to the condition of invariance under the charge conjugation; as a complex two component spinor which appears in general solution of that condition; or as a four component real bispinor obtained by taking the real and imaginary parts of the complex spinor, see, e.g., \cite{A1}, \cite{Aste}.  We prefer the latter form because it is in accordance with the fact that the set of real numbers is the proper algebraic number field for linear space of the Majorana bispinors \footnote{We prefer the convenient term `bispinor'  because there are four components, not to suggest that the Majorana bispinor is composed of two spinors. More precise term is `real spinor'.}.

There are several reasons for our interest in quantization of the classical Majorana field. 
Quantization of classical fields usually begins  with a Lagrangian, canonical momentum and Hamiltonian formulation, but in the  case of classical Majorana field there is a problem. Our preliminary investigation  shows that the classical massive Majorana field has a nonstandard, nonlocal Lagrangian, as opposed to  the pseudoclassical  Majorana field for which one can use essentially the Dirac Lagrangian.  Quantization of this latter field is rather straightforward because one can use the canonical quantization.   Quantization of the  classical Majorana field is less obvious, and this fact makes it interesting.  In the quantization presented below we do not need  classical Lagrangian at all. 

The main feature of the quantum theory of the Majorana field is well-known:  it predicts a relativistic, spin 1/2 particle, without an antiparticle \cite{Maj}. We would like to check whether one can arrive at this theory  starting from an expansion of the classical Majorana field into 
eigenfunctions of the so called axial momentum, a novel observable which has been introduced and discussed in our earlier papers   \cite{A2},  \cite{A3}. This expansion is distinguished by the fact that both the eigenfunctions and all coefficients  are real, like the field.   This is the second part of our motivation. 

There is also a third reason.  In literature, the quantum Majorana field often is obtained  by quantizing the complex Dirac field first,  and next imposing the condition of invariance under the charge conjugation, see, e. g., Section 6.2.3 in \cite{part1}.    Such a shortcut procedure, however, requires certain amount of care, as pointed out in  \cite{Fujikawa2} and \cite{Dvoeglazov}. Therefore, we would like to 
have a complete in itself  quantization in which only the Majorana field is considered. Such a self-contained approach seems to be missing in literature.

We consider only  the massive Majorana field.  When the field is  massless  a local gauge symmetry  is present, see Section 5.2 in \cite{A1}. In consequence, the structure of the quantum theory is very different from the massive case.  We shall not study it here in order to keep size of the paper within reasonable limits.

 It turns out that the presented below new path to the old result offers interesting
insights.  As the most important one we would consider the appearance of the real Clifford algebra generated by the so called  Majorana basis of Hermitian fermionic operators. This algebra seems to be the right mathematical structure for the quantized Majorana  field.  Another virtue of our approach is that it shows how one can construct the quantum theory  essentially by maintaining the relativistic invariance.   Last but not least, we see  that eigenfunctions of the axial momentum can successfully replace eigenfunctions of the ordinary momentum (i.e., the plane waves) in mode expansion for the Majorana field, and perhaps also for the Dirac field.

 The plan of our paper is as follows. In Section 2 we briefly recall the expansion of the classical Majorana field into eigenfunctions of the axial momentum \cite{A3}, for convenience of the reader and in order to fix notation. Section 3 is devoted to the quantization. The Majorana basis of operators is introduced, and generators of Poincar\'e transformations found. The Fock space and particle interpretation are 
 constructed in Section 4. In the 5th Section  we investigate the Majorana field operator in the Fock space.  Section 6 contains explicit construction of operators   representing in the Fock space  the space inversion (unitary  $\hat{\mbox{P}}$)  and the time reversal (antiunitary $\hat{\mbox{T}}$).   Summary and remarks are presented in Section 7. In  Appendix A we have collected useful formulas pertinent to the Lorentz boosts and the Wigner rotations.  Appendix B contains formulas  helpful in  computations of  commutators  in Sections 3 and 4.  
 
We use the Minkowski metrics $\eta = \mbox{diag}(1, -1,-1,-1)$. Throughout the paper summation over  repeated indices is understood.

\section{The classical Majorana field in the axial \\ momentum basis}

In this Section  we fix notation for the classical Majorana field, and we remind the axial momentum operator with its eigenfunctions called the axial plane waves. Next, we recall the expansion of the field in the basis of the axial plane waves presented in  \cite{A3}, and we adjust it for the present goal. The amplitudes $b^{\alpha}(\mathbf{p})$ introduced in this expansion are to be replaced by Hermitian operators in the process of quantization.  

 Let us begin with the Dirac equation  for the classical  Majorana field $\psi(x)$
 \begin{equation}
 i \gamma^{\mu} \partial_{\mu} \psi(x) - m \psi(x) =0,
 \end{equation}
where  $m >0$, and all matrices $\gamma^{\mu}$  are  purely imaginary. 
We use the following matrices
\vspace*{-0.1cm}
\[ \gamma^0 = \left(  \begin{array}{cc} 0 & \sigma_2 \\ \sigma_2 & 0 \end{array}  \right), \;\;    \gamma^1 = i \left(  \begin{array}{cc} - I_2 & 0 \\ 0 & I_2 \end{array}  \right), \;\;   \gamma^2 = i \left(  \begin{array}{cc} 0 & \sigma_1 \\ \sigma_1 & 0 \end{array}  \right),\]  \[  \gamma^3 = -i \left(  \begin{array}{cc} 0 & \sigma_3 \\ \sigma_3 & 0 \end{array}  \right),   \;\;\;\; \gamma_5= i \gamma^0 \gamma^1 \gamma^2 \gamma^3 = i \left(  \begin{array}{cc} 0 & \sigma_0 \\ - \sigma_0 & 0 \end{array}  \right).  \]
Here $\sigma_k$ are the Pauli matrices, and $\sigma_0$ is the two by two unit matrix.  All four components $\psi^{\alpha}$ of the Majorana field   are real valued functions on the Minkowski spacetime $M$.

The space of solutions of Eq.\ (1) is invariant with respect to the Poincar\'e  transformations 
\begin{equation} \psi'(x)= S(L) \psi(L^{-1}(x-a)), \end{equation} 
where the four-vector $a$  represents a translation in the Minkowski spacetime, and $L$ a proper ortochronous Lorentz transformation. 
The four by four real matrix $S(L)$ obeys the condition 
\[S(L)^{-1} \gamma^{\mu} S(L) = L^{\mu\;}_{\;\;\nu} \:\gamma^{\nu}.  \]
We shall consider also the discrete symmetries $\mbox{P}$ and $\mbox{T}$: 
\begin{equation}\psi_{\mbox{P}}(t, \mathbf{x}) = \eta_{\mbox{P}} i \gamma^0 \psi(t, - \mathbf{x}), \;\;\;  \psi_{\mbox{T}}(t, \mathbf{x}) = \eta_{\mbox{T}}  \gamma^0 \gamma_5 \psi(-t,  \mathbf{x}), \end{equation}
where the numerical factors $\eta_{\mbox{P}}, \eta_{\mbox{T}}$ are equal to 1 or -1. 

The axial momentum  operator  has the form \cite{A2} \[ \hat{\mathbf{p}}_5 =  -i \gamma_5 \nabla. \]  
Its normalized eigenfunctions, called the axial plane waves, read
\[
 \psi_{ \mathbf{p}}(\mathbf{x}) =  (2\pi)^{-3/2} \exp(i \gamma_5 \mathbf{p}\mathbf{x}) \: v,  
\]
where  $v$ an arbitrary real, constant and  normalized  bispinor,  $ v^T v=1$, $\:T$ denotes the matrix transposition.  The matrix $\exp(i \gamma_5 \mathbf{p}\mathbf{x})$  is real and orthogonal, because $\gamma_5^* = - \gamma_5$ and 
$\gamma_5^T = - \gamma_5$. Thus,
 \[ \hat{\mathbf{p}}_5 \psi_{ \mathbf{p}}(\mathbf{x}) =  \mathbf{p}\:  \psi_{ \mathbf{p}}(\mathbf{x}),  \;\; \int\!d^3x\:  \psi^{T}_{ \mathbf{p}}(\mathbf{x})\: \psi_{\mathbf{q}}(\mathbf{x}) = \delta(\mathbf{p} - \mathbf{q}). \]  Note that
 $  \exp(i \gamma_5 \mathbf{p}\mathbf{x}) =  \cos(\mathbf{p}\mathbf{x}) I_4  + i \gamma_5 \sin( \mathbf{p}  \mathbf{x} ). $
Detailed discussion  of rather interesting  properties of the axial momentum operator is given in \cite{A2} and \cite{A3}.

General solution of the Dirac equation (1) can be written as  time dependent superposition of the axial plane waves, see Section 4 in  \cite{A3},  
\begin{equation} \psi(\mathbf{x}, t) = \frac{1}{(2\pi)^{3/2}} \int\!\frac{d^3p}{E_p}\:\left( e^{i \gamma_5 (\mathbf{p} \mathbf{x} - E_p t)} v_{+}(\mathbf{p}) +   e^{-i \gamma_5 (\mathbf{p} \mathbf{x} - E_p t)}  v_{-}(\mathbf{p})  \right), \end{equation}
 where $E_p = \sqrt{\mathbf{p}^2 + m^2} >0$, and  $v_{\pm}(\mathbf{p})$ are real bispinors  which obey the following relations
 \[(E_p \gamma^0 - p^k \gamma^k) \gamma_5 v_{\pm}(\mathbf{p}) = \pm m  v_{\mp}(\mathbf{p}). \]
 Note that $\mbox{p}$ denotes eigenvalues of the axial momentum, not of the standard momentum. 
 
Next, we express the bispinor $v_{-}(\mathbf{p})$ by  $v_+(\mathbf{p})$, and     introduce a  basis $e_{\alpha}(\mathbf{p})$,  $\alpha=1,2,3, 4,$ in the linear space of  real bispinors $v_+(\mathbf{p})$,   \[v_+(\mathbf{p})= m \sqrt{E_p}\: e_{\alpha}(\mathbf{p}) \: b^{\alpha}(\mathbf{p}).\] 
 The coefficient $m \sqrt{E_p}$ is introduced in order to give the functions $b^{\alpha}(\mathbf{p})$ the dimension $cm^{3/2}$ consistent with the  postulated  anticommutators (8)  below.   We obtain 
\begin{equation}  
\psi(\mathbf{x}, t) = \frac{m}{(2\pi)^{3/2}} \int\!\frac{d^3p}{\sqrt{E_p}}\:\left[ e^{-i \gamma_5 p x} +  \frac{1}{m}  e^{i \gamma_5 px} p_{\mu}\gamma^{\mu}\gamma_5  \right] \:e_{\alpha}(\mathbf{p}) \: b^{\alpha}(\mathbf{p}), 
 \end{equation}
where  $p= (p^{\mu})=(E_p, p^1, p^2, p^3)^T,$ $\;px = E_p t -\mathbf{p} \mathbf{x},$ and $p_{\mu}\gamma^{\mu} = E_p \gamma^0 - p^k\gamma^k $. Note that because $\gamma_5$ anticommutes with $\gamma^{\mu}$, the order of matrices matters, e.g.,   $\exp(i\gamma_5 px)\: p_{\mu}\gamma^{\mu}  = p_{\mu}\gamma^{\mu}\: \exp(-i\gamma_5 px)$.

 The  basis $e_{\alpha}(\mathbf{p})$  is obtained  by  applying Lorentz boosts  $H(\mathbf{p})$ to the rescaled Cartesian basis  at  $\mathbf{p}=0$:  \[  e_{\alpha} ^{\eta}(\mathbf{0}) = \frac{\delta_{\alpha \eta}}{\sqrt{m}}; \;\;  e_{\alpha}(\mathbf{p}) =  S(H(\mathbf{p}))   e_{\alpha}(\mathbf{0}); \;\;  H(\mathbf{p}) \stackrel{(0)}{p}  = p, \]
 where $\stackrel{(0)}{p} = (m,0,0,0)^T$. The  index $\eta =1,2,3,4$ enumerates components  of the bispinor  $e_{\alpha}$. Bispinors are represented by matrices with one column and four rows. Explicit form of $H(\mathbf{p})$ can be found in the Appendix A.      The coefficient $1/\sqrt{m}$ is introduced in order to ensure that the basis bispinors $e_{\alpha}$ have the same dimensionality as $v_+$, which is $cm^{1/2}$.

The choice of the basis made above implies the following  Lorentz transformations  of the functions $b^{\alpha}$:   
\begin{equation} 
	b^{'\alpha}(\mathbf{p}) = \sqrt{\frac{E_{L^{-1}p }}{E_p} } \:  S({\cal R}(L,\mathbf{p}))_{\alpha\beta}\: b^{\beta}(l^{-1}(\mathbf{p})),   \end{equation} 
where  the matrix ${\cal R}(L,\mathbf{p}) = H^{-1}(\mathbf{p}) \: L\: H(l^{-1}(\mathbf{p}))$ leaves $\stackrel{(0)}{p}$ invariant. Therefore it represents a rotation, known as the Wigner rotation. The matrix $S({\cal R}(L,\mathbf{p}))$ is real and orthogonal, see the Appendix A.  The argument $ l^{-1}(\mathbf{p})$ of $b^{\beta}$ on the r.h.s. of formula (6)  is the spatial part of the four-vector $L^{-1}p$:  the $i$-th component of the three-vector $l^{-1}(\mathbf{p})$ is equal to $ (L^{-1})^{i\;}_{\;k} p^k + (L^{-1})^{i\;}_{\;0} E_p.$

In the case of translations,  $\psi'(x) = \psi(x-a).$ In consequence, \[b^{'\alpha}(\mathbf{p})  e_{\alpha}(\mathbf{p}) =   b^{\beta}(\mathbf{p})\: \exp(i \gamma_5 pa) \: e_{\beta}(\mathbf{p}), \] and
\begin{equation}  b^{'\alpha}(\mathbf{p}) =  \exp(i \gamma_5 pa)_{\alpha \beta}   b^{\beta} (\mathbf{p}) \end{equation} because \[ \exp(i \gamma_5 pa) \: e_{\beta}(\mathbf{p}) =  \exp(i \gamma_5 pa)_{\alpha \beta} \:e_{\alpha}(\mathbf{p}). \]
The matrix $\exp(i \gamma_5 pa)$ in (7) is real and orthogonal.

Real valued functions $ b^{\alpha}(\mathbf{p})$  parameterize the space of real solutions of the  Dirac  equation  (1)  for the classical Majorana field.  They are not restricted by any constraints.

 \section{Quantization: the algebraic part} 
 
 Here we consider  algebraic
 aspects of the would be quantum operators. Actual realization of them as Hermitian operators is possible only when we introduce a Hilbert space, e.g., the Fock space defined in the next Section. For convenience,  we shall use the term   `operator' already in the present Section.  
 
 The quantized Majorana field $\hat{\psi}(x)$  is obtained by replacing the real valued functions 
 $b^{\alpha}(\mathbf{p})$ with Hermitian operator valued functions $ \hat{b}^{\alpha}(\mathbf{p})$  \footnote{In fact, they turn out to be operator valued generalized functions of $\mathbf{p}$.}, which obey 
  the  anticommutation constraints \footnote{We anticipate the fermionic character of the field.}
 \[ [ \hat{b}^{\alpha}(\mathbf{p}), \; \hat{b}^{\beta}(\mathbf{q}) ]_+  = \kappa_{\alpha \beta}(\mathbf{p}) \:I  \:\delta(\mathbf{p} - \mathbf{q}), \]
 where $\kappa_{\alpha\beta}(\mathbf{p})$ are   real valued, dimensionless functions of  $\mathbf{p}$, $\kappa_{\alpha\beta}(\mathbf{p}) =\kappa_{\beta\alpha}(\mathbf{p})$,  and $I$ is the identity operator.  
We demand that these constraints  are invariant with respect to the Lorentz and translation transformations (6) and (7). Simple calculations show that in this case $ \kappa_{\alpha\beta}(\mathbf{p}) = \kappa_0 \:\delta_{\alpha\beta},$  where $\kappa_0 $ is a real, dimensionless, positive constant. Rescaling the operators $\hat{b}^{\alpha}$ we may put  $\kappa_0 =1$.  Thus, 
  \begin{equation}  [ \hat{b}^{\alpha}(\mathbf{p}), \; \hat{b}^{\beta}(\mathbf{q})  ]_+  = \delta_{\alpha\beta} \:I  \:\delta(\mathbf{p} - \mathbf{q}). \end{equation}
 Such a set of Hermitian fermionic operators  is known in condensed matter physics as the Majorana basis of operators.   
 
The crucial  point to be addressed is relativistic invariance  of the quantum theory. In such a theory   transformations (6) and (7) should be implemented by unitary operators
 $U(\Lambda, a)$, where $\Lambda \in SL(2, \mathbb{C})$. $SL(2, \mathbb{C})$ is the universal covering group of the proper ortochronous Lorentz group. The two groups are isomorphic in a vicinity of the unit element. Thus, $\Lambda$ can be regarded as a function of the Lorentz transformation $L$.  For a detailed discussion of the relation between these two groups  see, e.g., \cite{BLT}. Our task is to find the operators $U(\Lambda, a)$. To this end,  it suffices to consider infinitesimal transformations, i.e., from a small vicinity of the unit element. Then 
 \[ U(\Lambda(L), a) \cong  I + i a^{\mu} \hat{P}_{\mu} + \frac{i}{2} \omega^{\mu\nu} \hat{M}_{\mu\nu},\]
 where $a^{\mu}$ parameterize translations in the spacetime, and    $\omega_{\mu\nu}= - \omega_{\nu\mu}$ parameterize the proper orthochronous Lorentz group  around $I_4$, namely $L=\exp\omega$, where $\:\omega =(\omega^{\mu\;\;}_{\;\;\nu})$.  Because $\hat{M}_{\mu\nu} = -\hat{M}_{\nu\mu}$, there are six independent generators of Lorentz transformations.

   The Hermitian operators 
\begin{equation} \hat{P}_{\mu}= -i \left. \frac{\partial U(\sigma_0, a)}{\partial a^{\mu}}\right|_{a=0}, \; \hat{M}_{\mu\nu}= -i \left. \frac{\partial U(\Lambda(L), a=0)}{\partial \omega^{\mu\nu}}\right|_{\omega=0}, \end{equation} 
are identified with, respectively, the total four-momentum and the total angular momentum of the field -- the most important observables for the quantized field. The matrix $\sigma_0$  appears because it is the unit element in the $SL(2,\mathbb{C})$ group.  In the second formula (9) the matrix $\Lambda$ is regarded as function of the Lorentz matrix $L$. The operators 
$\hat{P}_{\mu}$, $\hat{M}_{\mu\nu}$ are called the generators of the representation  because they essentially determine the representation $ U(\Lambda(L), a)\;$  \cite{Raczka}.  

In the case of translations,  the postulated quantum version of condition (7) reads
\[ U^{-1}(\sigma_0,a) \:\hat{b}^{\alpha}(\mathbf{p}) \: U(\sigma_0, a) =   \exp(i\gamma_5 pa)_{\alpha\beta}    \hat{b}^{\beta} (\mathbf{p}). \] 
Differentiating both sides of this formula with respect to $a^{\mu}$ and putting $a=0$ we obtain the condition for $ \hat{P}_{\mu}$
\begin{equation} \left[ \hat{P}_{\mu}, \hat{b}^{\alpha}(\mathbf{p}) \right] = -p_{\mu} (\gamma_5)_{\alpha\beta}\: \hat{b}^{\beta}(\mathbf{p}),  \end{equation} where $p_0 = E_p =\sqrt{\mathbf{p}^2 + m^2}$. As shown in the Appendix B, there is a general formula for operators satisfying such  conditions. Condition (10) has the form of formula $(B1)$  with $\hat{P}_{\mu}$ and  
$ r_{\alpha\beta}(\mathbf{p}, \mathbf{q}) = i p_{\mu} (\gamma_5)_{\alpha\beta} \delta(\mathbf{p} - \mathbf{q}) $
in the place of $\hat{X}$ and $x_{\alpha\beta}(\mathbf{p}, \mathbf{q})$,   respectively. 
Formula $(B2)$ gives  the following Hermitian operators  
\begin{equation}  \hat{P}_{\mu} =   \frac{1}{2} \int d^3p \: p_{\mu}\: \hat{b}^{\alpha}(\mathbf{p})\: (\gamma_5)_{\alpha\beta}\: \hat{b}^{\beta}(\mathbf{p})  + d_{\mu} I,   \end{equation}   
where $d_{\mu}$ are arbitrary real constants.  
These operators commute with each other,   \[[ \hat{P}_{\mu},  \hat{P}_{\nu}] =0, \] as expected for the generators of translations. This can be checked with the help of formulas $(B3)$, $(B4)$.

Similar, but more tedious calculations give  the generators $\hat{M}_{\mu\nu}$.  
In order to obtain conditions for $\hat{M}_{\mu\nu}$ analogous to  (10)   we  postulate the quantum version of the Lorentz transformation (6)
\begin{equation}
U^{-1}(\Lambda(L), 0) \:\hat{b}^{\alpha}(\mathbf{p}) \:U(\Lambda(L), 0) = \sqrt{\frac{E_{L^{-1}p}  }{E_p} } \:  S({\cal R}(L,\mathbf{p}))_{\alpha\beta}\: \hat{b}^{\beta}(l^{-1}(\mathbf{p})). 
\end{equation}

We differentiate both sides of formula (12) with respect to $\omega_{jk}$  and next we put $\omega=0$. This   gives the following condition for $\hat{M}_{jk}$
\begin{equation}
[\hat{M}_{jk}, \hat{b}^{\alpha}(\mathbf{p}) ] =  \frac{i}{4} [\gamma^j, \gamma^k]_{\alpha\beta}\hat{b}^{\beta}(\mathbf{p}) - i \left(p^j \frac{\partial}{\partial p^k} - p^k \frac{\partial}{\partial p^j}          \right)  \hat{b}^{\alpha}(\mathbf{p}). 
\end{equation}
Calculations giving the r.h.s. are explained in the Appendix A. 
Formula (13) has the form as in $(B1)$ with 
\[ m^{jk}_{\alpha\beta}(\mathbf{p}, \mathbf{q}) =  \frac{1}{4} [\gamma^j, \gamma^k]_{\alpha\beta} \:\delta(\mathbf{p}- \mathbf{q}) -  \left(p^j \frac{\partial}{\partial p^k} - p^k \frac{\partial}{\partial p^j}  \right) \delta_{\alpha\beta} \:\delta(\mathbf{p}- \mathbf{q})        
\]
in place of  $x_{\alpha\beta}(\mathbf{p}, \mathbf{q}).$  Formula $(B2)$ gives
\begin{eqnarray}
\hat{M}_{jk} = \lefteqn{- \frac{i}{8} \int \! d^3p \: \hat{b}^{\alpha}(\mathbf{p}) [\gamma^j, \gamma^k]_{\alpha\beta}  \hat{b}^{\beta}(\mathbf{p}) }  \hspace*{2cm}     \\ &&  + \frac{i}{2} \int \! d^3p  \:\hat{b}^{\alpha}(\mathbf{p}) \left(p^j \frac{\partial}{\partial p^k} - p^k \frac{\partial}{\partial p^j}          \right)  \hat{b}^{\alpha}(\mathbf{p}) + d_{jk} I, \nonumber
\end{eqnarray}
where $d_{jk}= - d_{kj}$ are arbitrary real constants. 

Similarly, differentiation of formula (12) with respect to $\omega_{0k}$ at $\omega=0$ gives  the condition (see the Appendix A for details of the calculation of the r.h.s.)
\begin{eqnarray}
[\hat{M}_{0j}, \hat{b}^{\alpha}(\mathbf{p}) ] = \lefteqn{ - \frac{i}{4} \frac{1}{E_p +m} (p^r\delta_{sj} -  p^s \delta_{rj}) (\gamma^r \gamma^s)_{\alpha\beta}\hat{b}^{\beta}(\mathbf{p}) }\\ && + \frac{i}{2} \int\!d^3q \:  \left( (E_p \frac{\partial}{\partial p^j} - E_q \frac{\partial}{\partial q^j}) \delta(\mathbf{p} - \mathbf{q})    \right)  \hat{b}^{\alpha}(\mathbf{q}), \nonumber
\end{eqnarray}
which is satisfied by 
\begin{eqnarray}
	\hat{M}_{0j}  = \lefteqn{  \frac{i}{8} \int\!d^3p\:\frac{1}{E_p +m} (p^r\delta_{sj} -  p^s \delta_{rj}) \hat{b}^{\alpha}(\mathbf{p})(\gamma^r \gamma^s)_{\alpha\beta} \hat{b}^{\beta}(\mathbf{p}) }\\ && - \frac{i}{4} \int\!d^3p d^3q \: \hat{b}^{\alpha}(\mathbf{p}) \hat{b}^{\alpha}(\mathbf{q}) (E_p \frac{\partial}{\partial p^j} - E_q \frac{\partial}{\partial q^j}) \delta(\mathbf{p} - \mathbf{q})     +d_{0j}I,  \nonumber
\end{eqnarray}
where the real constants $d_{0j}$ are arbitrary. 

The generators should obey certain  commutator relations which follow directly from  the properties of the Poincar\'e group, see, e.g., \cite{Raczka}. In the  parameterization of the Poincar\'e group  introduced  above formula (9), they have the form 
\begin{equation}  [ \hat{P}_{\mu},  \hat{P}_{\nu}] =0, \end{equation}  
\begin{equation} [ \hat{M}_{\rho \lambda}, \hat{P}_{\mu} ] = i (\eta_{\mu\rho} \hat{P}_{\lambda} - \eta_{\mu\lambda} \hat{P}_{\rho}), \end{equation} 
\begin{equation} [ \hat{M}_{\alpha \beta}, \hat{M}_{\mu\nu} ] = i (\eta_{\alpha \mu} \hat{M}_{\beta \nu} - \eta_{\alpha \nu} \hat{M}_{\beta \mu}  -\eta_{\beta \mu} \hat{M}_{\alpha \nu} +\eta_{\beta \nu} \hat{M}_{\alpha \mu}),
\end{equation}
where $\eta_{\mu\rho}$ are components of the Minkowski metric tensor.

Our operators (11), (14) and (16) do not obey the 
commutation relations (18), (19) unless the constants $d_{\mu}, \: d_{\mu \nu}$ vanish.   To show this, we use formulas  $(B3)$, $(B4)$ from the Appendix B for the terms with the $\hat{b}^{\alpha}$ operators.  It turns out that already these terms alone satisfy the commutators (18), (19).   The terms with the identity operator of course give vanishing contributions to the l.h.s. of commutators (18) and (19), but they explicitly appear on the r.h.s's. In this way, we obtain from (18),  (19) the following conditions  
 \[ 0=  \eta_{\mu\rho} d_{\nu} - \eta_{\mu\nu} d_{\rho}, \;\;\; 
0 = \eta_{\alpha \mu} d_{\beta \nu} - \eta_{\alpha \nu} d_{\beta \mu}  +\eta_{\beta \nu} d_{\alpha \mu} -\eta_{\beta \mu} d_{\alpha \nu},\] which imply that $d_{\lambda} =0$ and $d_{\mu \nu}=0$. Nevertheless, we shall keep these  constants nonvanishing. The reason is that we  prefer the normal ordered generators in the Fock space, discussed in the next Section,
 because their eigenvalues are consistent with the particle interpretation. Indeed,  the normal ordered generators  are regarded as physical observables  for the quantized field,  in particular as its total four-momentum and total angular momentum.
The normal ordering can be interpreted as a special  choice of the constants   $d_{\mu}, \: d_{\mu \nu}$ \footnote{In this case the constants are given by integrals over $\mathbf{p}$ which  are divergent unless there is a cutoff.}.
 Thus,  because we insist on having the particle interpretation,  the relativistic invariance of the model  critically depends  on whether the normal ordered generators, and not the ones above,  obey the commutator relations (17), (18), and (19).  

Note that the Poincar\'e generators contain the imaginary unit $i$ as overall coefficient. Therefore the  operator $U(L(\Lambda), a)$ contains products of the operators $\hat{b}^{\alpha}(\mathbf{p})$ with only real coefficients.  Thus, the algebraic structure described above is a real, infinite dimensional Clifford algebra with the operators $\hat{b}^{\alpha}(\mathbf{p})$ as its Hermitian generating elements.

\section{The complex Fock space and  particle interpretation}

First, let us introduce two annihilation  and two creation operators  $\hat{a}_{\lambda}(\mathbf{p})$,  $ \hat{a}^{\dagger}_{\lambda}(\mathbf{p})$, where $\lambda=1, 2$, 
\begin{equation}
\hat{a}_{\lambda}(\mathbf{p})= c_{\lambda \alpha} \hat{b}^{\alpha}(\mathbf{p}), \;\;\; 	  \hat{a}^{\dagger}_{\lambda}(\mathbf{p}) = c^*_{\lambda \alpha} \hat{b}^{\alpha}(\mathbf{p}),
\end{equation}
where $c_{\lambda \alpha}$ are constants, $*$ denotes the complex conjugation, $\alpha =1,2,3,4.$  
 By assumption, these operators  have the following anticommutators
\begin{equation} \left[\hat{a}^{\dagger}_{\lambda}(\mathbf{p}), \: \hat{a}_{\sigma}(\mathbf{q})   \right]_+  = \delta_{\lambda \sigma} \delta(\mathbf{p} - \mathbf{q}) \: I,   \end{equation}  \begin{equation} \;\; \left[\hat{a}^{\dagger}_{\lambda}(\mathbf{p}), \: \hat{a}^{\dagger}_{\sigma}(\mathbf{q})   \right]_+  = 0 ,\:\;\;\;\;  \left[\hat{a}_{\lambda}(\mathbf{p}), \: \hat{a}_{\sigma}(\mathbf{q})   \right]_+  = 0.  \end{equation}
Inserting formulas (20) and using the anticommutators (8) we obtain from (21), (22)
constraints for the constants $c_{\lambda \alpha}$, namely 
\begin{equation}
c_{\lambda\alpha} c_{\sigma\alpha}=0, \;\;\; 	c^{}_{\lambda\alpha} c^*_{\sigma\alpha} = \delta_{\lambda \sigma}.
\end{equation}	
These constraints do not have unique solution because their l.h.s.'s are invariant 
with respect to arbitrary transformations of the form $c'_{\lambda \alpha} = c_{\lambda\beta} {\cal O}_{\beta \alpha}$, where ${\cal O}_{\beta\alpha}$ form a real, orthogonal,  four by four matrix.

With the Fock space and particle interpretation as the goal, we would like to obtain the four momentum operator in the form typical for quantum theory of free fields. It our case it reads 
\begin{equation} :\!\!\hat{P}_{\mu}\!\!: \;\;= \int\!\! d^3p\: p_{\mu}  \hat{a}_{\lambda}^{\dagger}(\mathbf{p})  \hat{a}_{\lambda}(\mathbf{p}),  \end{equation}
where $p_0 = \sqrt{\mathbf{p}^2 + m^2},$ $\;\lambda=1, 2$, and $:\; :$ denotes the normal ordering of products of the operators $\hat{a}_{\lambda}, \hat{a}^{\dagger}_{\lambda}$. 
We start from formula (11) for the operator $\hat{P}_{\mu}$, in which we substitute
\begin{equation}
\hat{b}^{\alpha}(\mathbf{p}) = c^*_{\lambda \alpha} \hat{a}_{\lambda}(\mathbf{p}) + c_{\lambda \alpha} 	 \hat{a}^{\dagger}_{\lambda}(\mathbf{p}).
\end{equation}
This formula is inverse to (20).  The constants $c_{\lambda \alpha} , c^*_{\lambda \alpha} $ obey conditions (23). In the resulting expression for $\hat{P}_{\mu}$ there are unwanted terms 
with the products $ \hat{a}_{\lambda} \hat{a}_{\sigma}$ and $\hat{a}^{\dagger}_{\lambda} \:\hat{a}^{\dagger}_{\sigma}$.  These terms vanish when the constants $c_{\lambda\alpha}$ obey the following conditions
\begin{equation}
	c_{\lambda \alpha}\: (\gamma_5)_{\alpha \beta}\: c_{\sigma \beta} =0.
	\end{equation}
If we add yet another condition, namely  
 \begin{equation}
 	c_{\lambda \alpha} (\gamma_5)_{\alpha \beta} c^*_{\sigma \beta} = \delta_{\lambda\sigma}, 
 \end{equation}
the four-momentum  operator acquires the form  
\[ \hat{P}_{\mu} 
= \frac{1}{2}\! \int\!\! d^3p\: p_{\mu} \left( \hat{a}_{\lambda}^{\dagger}(\mathbf{p})  \hat{a}_{\lambda}(\mathbf{p}) -  \hat{a}_{\lambda}(\mathbf{p}) \hat{a}_{\lambda}^{\dagger}(\mathbf{p}) \right) +d_{\mu} I. \]
In the last step  we apply the normal ordering and drop the term $d_{\mu}I$. On a heuristic level, one can say that the term generated by the normal ordering, which is proportional to $I$, is canceled by the term $d_{\mu} I$ with appropriately chosen constant $d_{\mu}$.  The final form of the four-momentum operator is given by formula (24).

At this point it is clear that we may use the  standard   Fock space as the complex Hilbert space for the quantized Majorana field.  In particular, the Fock vacuum state $|0\rangle$ is defined by the conditions
\begin{equation} \hat{a}_{1}(\mathbf{p}) |0\rangle =0, \;\; \hat{a}_{2}(\mathbf{p}) |0\rangle =0. \end{equation}
The Fock basis of quantum states  of the field is created by the operators $ \hat{a}_{\lambda}^{\dagger}(\mathbf{p})$ acting on the vacuum state.  Such states are eigenvectors of the $ :\!\!\hat{P}_{\mu}\!\!:$ operators.  Note that the Fock space is linear over the set of complex numbers $\mathbb{C}$, while the classical Majorana field is real valued.

We see that the energy operator $:\!\!\hat{P}_0\!\!:$  is non negative. Let us recall that when quantizing the Dirac field,  at certain stage one has to redefine  the vacuum state:  an empty vacuum defined by conditions analogous to (28) is abandoned in favor of the state known as the Dirac sea, otherwise the energy operator is not bounded from below. In the Majorana case, such redefinition is not needed.  In this sense, the Dirac sea is absent  here.

Condition (27) is satisfied  if $c^*_{\sigma \beta}$ are components of two orthonormal eigenspinors of the matrix $\gamma_5$  corresponding to eigenvalue $+1$. Thus,
\[ (\gamma_5)_{\alpha \beta} c^*_{\sigma \beta} = c^*_{\sigma \alpha}\]
($\sigma=1,2$ enumerates the eigenvectors).
 Because
$\gamma_5$ is imaginary, $c_{\sigma \beta}$ give other two eigenspinors which correspond to the eigenvalue $-1$.
Condition (26) is then reduced to $ 	c_{\lambda \alpha}\:  c_{\sigma \alpha} =0$, which coincides with the first condition in (23). Also the second condition in (23) is  
satisfied if the eigenspinors are orthonormal. The eigenvalues of $\gamma_5$ are 
double degenerate, hence the orthonormal  eigenspinors are not fixed uniquely.
We take 
\[ (c^*_{1\: \alpha}) = \frac{1}{\sqrt{2}} \left( \begin{array}{c} i  \\ 0 \\ 1 \\0  \end{array}  \right), \;\;\; (c^*_{2\: \alpha}) = \frac{1}{\sqrt{2}} \left(\begin{array}{c} 0  \\ i \\ 0 \\1 \end{array}  \right).  \]
The resulting relation between the operators $\hat{b}^{\alpha}$ and $\hat{a}_{\lambda}, \: \hat{a}^{\dagger}_{\lambda}$ can be summarized in the matrix form
\begin{equation}
 \left( \begin{array}{c} \hat{b}^1(\mathbf{p})   \\ \hat{b}^2(\mathbf{p}) \\ \hat{b}^3(\mathbf{p}) \\\hat{b}^4(\mathbf{p})  \end{array}  \right)	=  \frac{1}{\sqrt{2}} \left( \begin{array}{rrrr} i & 0&-i&0   \\ 0&i&0&-i \\ 1&0&1&0 \\ 0&1&0&1  \end{array}  \right)  \left( \begin{array}{l} \hat{a}_{1}(\mathbf{p})   \\ \hat{a}_{2}(\mathbf{p}) \\ \hat{a}^{\dagger}_{1}(\mathbf{p}) \\\hat{a}^{\dagger}_{2}(\mathbf{p})  \end{array}  \right)
\end{equation}
The matrix on the r.h.s. of formula (29) (with the factor $1/\sqrt{2}$ included) is unitary.

The three generators of  spatial rotations,  $:\!\hat{M}_{12}\!:, \: :\!\hat{M}_{23}\!:$, and $:\!\hat{M}_{31}\!:$, are obtained from formula (14). We use formula (29) for $\hat{b}^{\alpha}(\mathbf{p})$, next we apply the normal ordering, and remove by hand the terms proportional to the identity operator $I$. It turns out that the spin part comes out in a nonstandard form, namely $ \hat{M}_{12} \supset \sigma_1, \hat{M}_{23} \supset \sigma_2, \hat{M}_{31} \supset \sigma_3.$  Therefore we apply additional unitary transformation,
\begin{equation} \left( \begin{array}{c} \hat{a}_1(\mathbf{p}) \\ \hat{a}_2(\mathbf{p}) \end{array} \right) = Q \left(\begin{array}{c} \hat{d}_1(\mathbf{p}) \\ \hat{d}_2(\mathbf{p}) \end{array} \right),   \end{equation}
with the matrix $Q$ 
\[ Q= \frac{1}{\sqrt{2}} \left(\begin{array}{rr} 1 & -i \\ 1 & i \end{array} \right).  \] 
This transformation cyclically permutes the Pauli matrices \footnote{It follows that $Q^3$ commutes with the Pauli matrices. Indeed, $Q^3 = \exp(i \pi/4)\sigma_0.$}: 
\[ Q^{\dagger} \sigma_1 Q = \sigma_3,  \:Q^{\dagger} \sigma_2 Q = \sigma_1,\: Q^{\dagger} \sigma_3 Q = \sigma_2.      \]
Anticommutators of the operators $ \hat{d}_{\lambda}(\mathbf{p}),\: \hat{d}^{\dagger}_{\lambda}(\mathbf{p})$  have the form (21),  (22) of course, 
\[ \left[\hat{d}^{\dagger}_{\lambda}(\mathbf{p}), \: \hat{d}_{\sigma}(\mathbf{q})   \right]_+  = \delta_{\lambda \sigma} \delta(\mathbf{p} - \mathbf{q}) \: I, \] \[ \left[\hat{d}^{\dagger}_{\lambda}(\mathbf{p}), \: \hat{d}^{\dagger}_{\sigma}(\mathbf{q})   \right]_+  = 0 ,\:\;\;\;\;  \left[\hat{d}_{\lambda}(\mathbf{p}), \: \hat{d}_{\sigma}(\mathbf{q})   \right]_+  = 0.  \]
The definition (28) of the  vacuum state can be  equivalently written as 
\[ \hat{d}_{\lambda}(\mathbf{p}) |0\rangle =0. \]
Transformations (29) and (30) together give the following unitary transformation
\begin{equation}
	\left( \begin{array}{c} \hat{b}^1(\mathbf{p})   \\ \hat{b}^2(\mathbf{p}) \\ \hat{b}^3(\mathbf{p}) \\\hat{b}^4(\mathbf{p})  \end{array}  \right)	=  \frac{1}{2} \left( \begin{array}{rrrr} i & 1&-i&1   \\ i&-1&-i&-1 \\ 1&-i&1&i \\ 1&i&1&-i  \end{array}  \right)  \left( \begin{array}{l} \hat{d}_{1}(\mathbf{p})   \\ \hat{d}_{2}(\mathbf{p}) \\ \hat{d}^{\dagger}_{1}(\mathbf{p}) \\\hat{d}^{\dagger}_{2}(\mathbf{p})  \end{array}  \right).
\end{equation}
Transformation (30)  applied in the four-momentum operator gives 
\[ :\!\!\hat{P}_{\mu}\!\!: \;\;= \int\!\! d^3p\: p_{\mu}  \hat{d}_{\lambda}^{\dagger}(\mathbf{p})  \hat{d}_{\lambda}(\mathbf{p}). \]  

 The final result for the generators $\hat{M}_{ik}$ reads
\begin{equation}
	:\!\hat{M}_{ik}\!: = \epsilon_{ikj}  \int \! d^3p\: \hat{d}_{\lambda}^{\:\dagger}(\mathbf{p}) \left( -\frac{1}{2}(\sigma_j)_{\lambda \eta}  +i \delta_{\lambda \eta} \: \epsilon_{jmn} p^m \frac{\partial}{\partial p^n}  \right)\hat{d}_{\eta}(\mathbf{p}), 
\end{equation}	
Here $(\sigma_j)_{\lambda \eta} $ denotes matrix elements of the Pauli matrices, and  $\epsilon_{ikl}$ is the totally antisymmetric symbol, $\epsilon_{123}=+1$.  We notice  the spin part with the Pauli matrices,   and the angular momentum part with the derivatives  $\partial/ \partial p^n$.

The generators of boosts $:\!\hat{M}_{0k}\!:$ are obtained from formula (16) in a similar  manner, 
\begin{equation}
	:\!\hat{M}_{0k}\!: =  \int \! d^3p\: \hat{d}_{\lambda}^{\:\dagger}(\mathbf{p}) \left( \frac{\epsilon_{kjl}\: p^l}{2 (E_p +m)} (\sigma_j)_{\lambda \eta}  -i  (E_p  \frac{\partial}{\partial p^k} +\frac{p^k}{2E_p})\delta_{\lambda \eta} \right)\hat{d}_{\eta}(\mathbf{p}). 
\end{equation}

The normal ordered generators satisfy commutator relations (17), (18) and (19). This can be checked with the help of formula (B5)  from the Appendix B. 
We conclude that the constructed quantum model in the Fock space is  relativistically invariant. 

The index $\lambda = 1, 2$ can be related to the eigenvalues of the operator $\hat{\Sigma}^3,$ which is the 3rd component of the spin operator
\[ \hat{\Sigma}^i = \frac{1}{2} \int \! d^3p\: \hat{d}_{\kappa}^{\:\dagger}(\mathbf{p})  (\sigma_i)_{\kappa \eta}  \hat{d}_{\eta}(\mathbf{p}), \]  
in the single particle subspace of the Fock space. Such subspace is spanned on the basis states \[ | \mathbf{p}\: \lambda  \rangle = \hat{d}_{\lambda}^{\dagger}(\mathbf{p}) 
|0 \rangle, \]
which are eigenstates of $:\!\!\hat{P}^{\mu}\!\!: $ and $\hat{\Sigma}^3$:
\[:\!\!\hat{P}^{\mu}\!\!:  | \mathbf{p}\: \lambda  \rangle = p^{\mu} | \mathbf{p} \:\lambda  \rangle,  \;\;\; \hat{\Sigma}^3 | \mathbf{p} \;1  \rangle = \frac{1}{2}\: | \mathbf{p} \;1  \rangle, \;\;\; \hat{\Sigma}^3 | \mathbf{p} \; 2  \rangle = - \frac{1}{2} \:| \mathbf{p} \;2  \rangle \]
where $p^0 = \sqrt{\mathbf{p}^2 + m^2}$. 

General single particle state has the form $|\psi\rangle = \int\!\! d^3p \: \psi_{\lambda}(\mathbf{p})\: |\mathbf{p} \lambda \rangle$.  Such states form a subspace of the Fock space which is  invariant under 
 the Poincar\'e transformations generated by the operators $ :\!\!\hat{P}_{\mu}\!\!:$, 	$\;:\!\hat{M}_{ik}\!:$, and $:\!\hat{M}_{0k}\!:$. In fact, we have  obtained    the unitary irreducible representation of the Poincar\'e group in the single particle subspace of the Fock space. It is  characterised by  spin 1/2, the time-like four momentum    $p_{\mu}p^{\mu} =m^2>0$, and the positive energy $E_p>0$.

\section{The field operator in the Fock space }

The quantum field operator in the Fock space is obtained  from formula (5) by replacing 
the classical variables $b^{\alpha}(\mathbf{p})$ with the operators  $\hat{b}^{\alpha}(\mathbf{p})$, 
\begin{equation}  
	\hat{\psi}(\mathbf{x}, t) = \frac{m}{(2\pi)^{3/2}} \int\!\frac{d^3p}{\sqrt{E_p}}\:\left[ e^{-i \gamma_5 p x} +  \frac{1}{m}  e^{i \gamma_5 px} p_{\mu}\gamma^{\mu}\gamma_5  \right] \:e_{\alpha}(\mathbf{p}) \: \hat{b}^{\alpha}(\mathbf{p}), 
\end{equation}
where $\hat{b}^{\alpha}(\mathbf{p})$ are to be eliminated with the help of relation (31). 

This form of the field operator 
can be significantly modified. First, using the definition of the basis $e_{\alpha}(\mathbf{p})$ and  formula $ S(H(\mathbf{p}))^{-1} p_{\mu} \gamma^{\mu} S(H(\mathbf{p})) = m \gamma^0$, we may rewrite it as
\begin{eqnarray*}\hat{\psi}(\mathbf{x}, t) =  \lefteqn{ \frac{m}{(2\pi)^{3/2}} \int\!\frac{d^3p}{\sqrt{E_p}} \:  S(H(\mathbf{p})) \:\left[ e^{i \gamma_5 (\mathbf{p} \mathbf{x} - E_p t)}\right.} & \\ & \left. \;\;\;\;\; \;\;\;\;\;\;\;\;\;\;\;\; \;\;\;\;\;\;\;\;\;\;\;\; +    e^{-i \gamma_5 (\mathbf{p} \mathbf{x} - E_p t)}  \gamma^0 \gamma_5  \right] \:e_{\alpha}(\mathbf{0}) \: \hat{b}^{\alpha}(\mathbf{p}). \end{eqnarray*}
 Next, we  substitute  $\exp(\pm i \gamma_5 px) = \cos(px) I_4 \pm i\gamma_5 \sin(px)$, and   notice that the matrices  $\;i \gamma_5$,  $i \gamma^0$, and  
$\gamma^0 \gamma_5 = \left(  \begin{array}{rrrr} 0 &-1 &0&0 \\ 1&0&0&0 \\ 0&0& 0&1\\ 0&0&-1&0  \end{array}  \right)$  merely permute the basis bispinors $e_{\alpha}(\mathbf{0})$ and, in some cases, change their sign.  The $\sin$ and $\cos$ functions are expressed  by $\exp(\pm i pa)$ functions. Finally,  we introduce the creation and annihilation operators using formula (31).  After all these steps we obtain  the field operator in a more transparent form,
\begin{equation}   \hat{\psi}(\mathbf{x}, t) =\sqrt{m}  \int\!\frac{d^3p}{\sqrt{ (2\pi)^3E_p}} \left[ e^{- i px}\:  v_{\lambda}(\mathbf{p}) \: \hat{d}_{\lambda}(\mathbf{p})     +    e^{ i px} \: v^*_{\lambda}(\mathbf{p}) \: \hat{d}^{\:\dagger}_{\lambda}(\mathbf{p})  \right],  \end{equation}
where $ v_{1}(\mathbf{p})  = S(H(\mathbf{p}))v_1$,  $\;     v_{2}(\mathbf{p})=S(H(\mathbf{p}))v_2, $  and
 \[v_{1}= \frac{1}{\sqrt{2}}  \:\left(\begin{array}{r} 0\\ i \\ 1 \\ 0 \end{array} \right), \;\; \; v_{2}=\frac{1}{\sqrt{2}}  \:\left(\begin{array}{r} 1 \\ 0 \\ 0 \\ i \end{array} \right). \]
 The constant bispinors  $v_{\lambda}, v^*_{\lambda}$  are normalized eigenvectors of  the matrix  $\gamma^0$,  
 \begin{equation} \gamma^0 v_{\lambda} =  v_{\lambda}, \;\;\; \gamma^0 v^*_{\lambda} = - v^*_{\lambda}.  \end{equation}  
The definition of the boosted bispinors $v_{1}(\mathbf{p}), v_{2}(\mathbf{p})$ implies that
\[ (m\gamma^0 + p^i \gamma^0 \gamma^i) v_{\lambda}(\mathbf{p}) = E_p \: v_{\lambda}(\mathbf{p}), \;\;\;  \overline{v_{\lambda}(\mathbf{p})} v_{\sigma}(\mathbf{p}) = \delta_{\lambda \sigma}, \]
where $ \overline{v_{\lambda}(\mathbf{p})} =  v_{\lambda}^{\dagger}(\mathbf{p}) \gamma^0 $.
The first equation shows that they are eigenvectors of the Hermitian matrix $ m\gamma^0 + p^i \gamma^0 \gamma^i$ (which coincides with the Dirac Hamiltonian), while  the second one  
gives the normalization of the boosted bispinors. For the boosted complex conjugate bispinors,  $ v^*_{1}(\mathbf{p})  = S(H(\mathbf{p}))v^*_1\:$ and  $\:     v^*_{2}(\mathbf{p})=S(H(\mathbf{p}))v^*_2 $,  we have
 \[ (m\gamma^0 + p^i \gamma^0 \gamma^i) v^*_{\lambda}(-\mathbf{p}) = -E_p \: v^*_{\lambda}(-\mathbf{p}), \;\;\;  \overline{v^*_{\lambda}(\mathbf{p})} v^*_{\sigma}(\mathbf{p}) = -\delta_{\lambda \sigma}. \]

The quantized Majorana field  obeys the Dirac equation. First, using formulas (35) and (36) we  write 
\[ m \hat{\psi}(\mathbf{x}, t) =  \int\!\frac{d^3p\: \sqrt{m}}{\sqrt{ (2\pi)^3E_p}} \:  S(H(\mathbf{p})) \: \left[ e^{- i px}\: m\gamma^0 v_{\lambda} \: \hat{d}_{\lambda}(\mathbf{p})      -    e^{ i px} \: m\gamma^0 v^*_{\lambda} \: \hat{d}^{\:\dagger}_{\lambda}(\mathbf{p})  \right]. \]
Next,   we apply on the r.h.s. the formula
$  S(H(\mathbf{p}))m \gamma^0 S(H(\mathbf{p}))^{-1} =  p_{\mu} \gamma^{\mu}$, 
\begin{eqnarray*}
 m \hat{\psi}(\mathbf{x}, t) =   \int\!\frac{d^3p\:\sqrt{m}}{\sqrt{ (2\pi)^3E_p}} \:  \: \left[ e^{- i px}\:  p_{\mu} \gamma^{\mu}  S(H(\mathbf{p}))v_{\lambda} \: \hat{d}_{\lambda}(\mathbf{p})  \right. \;\;\;\; \;\;\\  \;\;\;\;\;\;\;\;\;\;  \left.   -    e^{ i px} \:  p_{\mu} \gamma^{\mu}  S(H(\mathbf{p}))v^*_{\lambda} \: \hat{d}^{\:\dagger}_{\lambda}(\mathbf{p})  \right]= \:i \gamma^{\mu} \partial_{\mu} \hat{\psi}(x).   \end{eqnarray*}

The  field $\hat{\psi}(\mathbf{x}, t)$ obeys also the following equation
\begin{equation}
\partial_t \hat{\psi}(\mathbf{x}, t) = i [ :\!\!\hat{P}_0\!\!: , \:\hat{\psi}(\mathbf{x}, t)].	
\end{equation}	
It has the form of evolution equation for quantum operators in the Heisenberg picture. 
For this reason, we may consider $:\!\!\hat{P}_0\!\!:$ as the Hamiltonian of the quantized Majorana field, and the time dependent field  $\hat{\psi}(\mathbf{x}, t)$ as the field operator in the Heisenberg picture. It is important conceptual step. The point is that the presented above quantization is not based on canonical formalism, in particular, we have not considered any classical Hamiltonian. The operator   $:\!\!\hat{P}_0\!\!:$ has been introduced in Section 4 as the generator of time translations in the context of representations of the Poincar\'e group, not  as a quantum version of certain classical Hamiltonian. 

The field operator (35) is local in the sense proper for fermionic fields: bilinear local operators of the  form $\hat{\psi}^{\alpha}(x) A_{\alpha \beta}\: \hat{\psi}^{\beta}(x)$,  $\;\hat{\psi}^{\alpha}(y) B_{\alpha \beta}\: \hat{\psi}^{\beta}(y)$, where $A_{\alpha \beta},  B_{\alpha \beta}$ are 
complex numbers, commute if $(x-y)^2 <0$.  The commutator of such operators is proportional to the anticommutator $[\hat{\psi}^{\alpha}(x),  \;\hat{\psi}^{\beta}(y)]_+$  which can be calculated easily, 
\begin{equation}  \left[\hat{\psi}^{\alpha}(x),  \;\hat{\psi}^{\beta}(y)\right]_+ = ( (im I_4 - \gamma^{\mu} \frac{\partial }{\partial x^{\mu}})\gamma^0)_{\alpha \beta} \Delta(x-y), \end{equation}
where $\Delta(x-y)$ is the Jordan-Pauli function, $\Delta(x-y)=0$ if $(x-y)^2 <0.$ 

Let us close this Section with a remark on the  well-known  current density $j^{\mu}(x) =  \overline{\psi}(x) \gamma^{\mu} \psi(x)$, where $\psi$ is the classical Majorana field, $\overline{\psi}(x) = \psi(x)^T \gamma^0$.  Its existence poses a puzzle.  Because the current is conserved, $\partial_{\mu} j^{\mu}=0$
if $\psi$ is a solution of the Dirac equation (1), there exists the conserved charge 
$Q=\int \! d^3x \: j^0(\mathbf{x}, t)$. Usually, such a conserved charge  is associated with a $U(1)$ symmetry, or rather  $SO(2)$ because the field is real -- we invoke here the inverse Noether's theorem. However, in the Majorana case it is hard to point out such a symmetry. Analyzing this problem, the first question is about existence of Lagrangian for the classical real valued Majorana field, because a  Lagrangian is needed in the Noether theorem. Rather surprisingly, it turns out that the answer is not quite trivial.  Because this topic clearly lies far outside the scope of the present paper,  we will not pursue it here. (We plan  a separate manuscript devoted to it.) Instead, let us consider  the quantum counterpart of the current density $j^{\mu}(x)$. In order to write it, one has to regularize the product of field operators  preserving Hermiticity of the current. We choose the point-splitting regularization,
\[ \hat{j}_{\epsilon}^{\mu}(x) = \frac{1}{2} \left(\hat{\psi}(x)^T \gamma^0 \gamma^{\mu} 
\hat{\psi}(x+\epsilon)  + \hat{\psi}(x+\epsilon)^T \gamma^0 \gamma^{\mu} 
\hat{\psi}(x)\right), \]
where the four-vector $\epsilon$ is constant and non-vanishing. Because all matrices $\gamma^0\gamma^{\mu}$ are  symmetric, 
\begin{eqnarray*} \hat{j}_{\epsilon}^{\mu}(x) =\lefteqn{  \frac{1}{2} (\gamma^0\gamma^{\mu})_{\beta\alpha} \left[ \psi^{\alpha}(x), \: \psi^{\beta}(x+\epsilon)  \right]_+} & \\ &\;\;\;\;   = \frac{1}{2} Tr(\gamma^0\gamma^{\mu}  (im I_4 - \gamma^{\nu} \frac{\partial }{\partial x^{\nu}})\gamma^0) \Delta(-\epsilon) =0.  \end{eqnarray*}
The last equality follows from $Tr \:\gamma^{\mu}=0$  and $ \partial \Delta(-\epsilon) / \partial {x^{\nu}} =0$. 
This result for the regularized current suggests that the current and the related charge may not exist in the quantum theory of the Majorana field.

\section{The discrete symmetries $\mbox{P}$ and $\mbox{T}$}

\subsection{The space inversion $\mbox{P}$}
	
As mentioned in Section 2, the space of solutions of the Dirac equation (1) is invariant with respect to the transformation
\begin{equation}
	\psi_{\mbox{P}}(\mathbf{x},t) =  \eta_{\mbox{P}} i \gamma^0 \psi(-\mathbf{x},t), 
	\end{equation}
which represents the space inversion $\mbox{P}: \mathbf{x} \rightarrow  - \mathbf{x}$. The coefficient $\eta_{\mbox{P}}$ is real. 
Note that $ 	(\psi_{\mbox{P}})_{\mbox{P}}(\mathbf{x},t) = - \eta_{\mbox{P}}^2 \:\psi(\mathbf{x},t)$. 
Because $\mbox{P}^2=I$, we expect that $	(\psi_{\mbox{P}})_{\mbox{P}}$ is physically equivalent to $\psi$. This  is the case when $\eta_{\mbox{P}}^2=1$ \footnote{It a well-known fact that the overall sign of  fermionic fields is physically irrelevant.}.   
Inserting formula (5) for the Majorana field  in  (39) and using the definition of the basis bispinors $e_{\alpha}(\mathbf{p})$ we obtain the corresponding transformation 
of the functions $b^{\alpha}(\mathbf{p})$: 
\begin{equation}
b^{\alpha}_{\mbox{P}}(\mathbf{p}) =  \eta_{\mbox{P}} i (\gamma_5)_{\alpha \beta} b^{\beta}(- \mathbf{p}).	
\end{equation}

The space inversion can be implemented in the quantum theory by constructing a unitary operator $\hat{\mbox{P}}$ in the Fock space such that 
\begin{equation}
\hat{\mbox{P}}\: \hat{b}^{\alpha}(\mathbf{p}) \hat{\mbox{P}}^{-1} =   \eta_{\mbox{P}} i (\gamma_5)_{\alpha \beta} \hat{b}^{\beta}(- \mathbf{p}).	
\end{equation}

Formula (41) and transformation (31) give 
\begin{equation}
	\hat{\mbox{P}}\: \hat{d}_{\lambda}(\mathbf{p}) \hat{\mbox{P}}^{-1} =   \eta_{\mbox{P}}\: i \: \hat{d}_{\lambda}(- \mathbf{p}), \;\;\; \hat{\mbox{P}}\: \hat{d}^{\dagger}_{\lambda}(\mathbf{p}) \hat{\mbox{P}}^{-1} =  - \eta_{\mbox{P}}\: i \: \hat{d}^{\dagger}_{\lambda}(- \mathbf{p}).
	\end{equation}

The second formula in (42)  determines the linear operator $\hat{\mbox{P}}$ provided that we know how it acts on the vacuum state. We assume that the vacuum state is invariant \footnote{Discussion of this point within the general framework of relativistic quantum theory of fields can be found in Section 3.4 of Ref. \cite{BLT}.} 
\begin{equation} \hat{\mbox{P}} |0 \rangle = |0\rangle. \end{equation}
This assumption is consistent with the first formula in (42). (It would be inconsistent if, for example, on the r.h.s. of that formula there was the operator  $\hat{d}^{\dagger}_{\lambda}$.)
The basis in the n-particle sector of the Fock space is formed by vectors
\begin{equation} |\mathbf{p}_1 \lambda_1,\: \mathbf{p}_2 \lambda_2, \dots \mathbf{p}_n \lambda_n \rangle = \frac{1}{\sqrt{n!}} \hat{d}^{\dagger}_{\lambda_1}(\mathbf{p}_1) \hat{d}^{\dagger}_{\lambda_2}(\mathbf{p}_2), \dots \hat{d}^{\dagger}_{\lambda_n}(\mathbf{p}_n)  |0\rangle .
\end{equation}
It is clear that 
\begin{equation}\hat{\mbox{P}}\:|\mathbf{p}_1 \lambda_1,\: \mathbf{p}_2 \lambda_2, \dots \mathbf{p}_n \lambda_n \rangle = (-\eta_{\mbox{P}} i)^n \:|-\mathbf{p}_1 \lambda_1,\: -\mathbf{p}_2 \lambda_2, \dots -\mathbf{p}_n \lambda_n \rangle \end{equation}
Note that the  $|(-\eta_{\mbox{P}} i)^n|=1$, therefore transformation (45) does not change the norm of the basis vectors. 
Formulas (43) and (45) are taken for the definition of the  operator $\hat{\mbox{P}}$ in the Fock space. In order to calculate its action on arbitrary state it is sufficient to expand that state  in the basis (44). Formulas (42), as well as unitarity of $\hat{\mbox{P}}$, now reappear as easy to prove theorems.

The four-momentum operator $:\! \hat{P}_{\mu}\!:$ has the standard  four-vector transformation law  with respect to the space inversion, namely 
\[ \hat{\mbox{P}}\: :\!\! \hat{P}^{0}\!\!: \:\hat{\mbox{P}}^{-1} = \: :\!\! \hat{P}^{0}\!\!:, \;\;\; \hat{\mbox{P}}\: :\!\! \hat{P}^{i}\!:\! \:\hat{\mbox{P}}^{-1} = - :\!\! \hat{P}^{i}\!:\! . \]

Transformation law of the field operator $\hat{\psi}$ mimics  formula (39) for the classical field,
\[  \hat{\mbox{P}} \hat{\psi}(\mathbf{x},t) \hat{\mbox{P}}^{-1} = \eta_{\mbox{P}} i \gamma^0  \hat{\psi}(-\mathbf{x},t). 
\]
 To see this, compute  $ \hat{\mbox{P}} \hat{\psi}(\mathbf{x},t) \hat{\mbox{P}}^{-1}$ using formulas  (35) and (42). Next, recover the matrix $\gamma^0$  with the help of Eqs.\ (36), and  move it to the left using formula $S(H(\mathbf{p})) \gamma^0 = \gamma^0 S(H(-\mathbf{p}))$. In
  the last step change the integration variable from $\mathbf{p}$ to $-\mathbf{p}$.

\subsection{The time reversal $\mbox{T}$}
 
 Time reversal $\mbox{T}$ acts on the classical field as follows
 \begin{equation}
 \psi_{\mbox{T}}(\mathbf{x},t) = \eta_{\mbox{T}} \gamma^0\gamma_5  \psi(\mathbf{x},-t).
 \end{equation}
It is a symmetry of the Dirac Eq.\ (1) in the sense that if $ \psi(\mathbf{x},t)$ is a solution of it, so is $\psi_{\mbox{T}}(\mathbf{x},t)$. The coefficient $\eta_{\mbox{T}}$ is real.
Similarly as in the case of space inversion, $\eta_{\mbox{T}}^2 =1$ and $(\psi_{\mbox{T}})_{\mbox{T}}(\mathbf{x},t)= - \psi_{\mbox{T}}(\mathbf{x},t)$. Calculations analogous the ones leading to formula (40)  give
\[  b^{\alpha}_{\mbox{T}}(\mathbf{p}) = \eta_{\mbox{T}} (\gamma^0 \gamma_5)_{\alpha \beta}  b^{\beta}(-\mathbf{p}).
\]
In order to implement this transformation in the Fock space we seek a unitary, or antiunitary, operator $\hat{\mbox{T}}$   such that
\begin{equation}
	\hat{\mbox{T}}\: \hat{b}^{\alpha}(\mathbf{p}) \hat{\mbox{T}}^{-1} =   \eta_{\mbox{T}} i (\gamma^0\gamma_5)_{\alpha \beta} \hat{b}^{\beta}(- \mathbf{p}).	
\end{equation}
Moreover, it should leave the vacuum state unchanged, 
\begin{equation}
	\hat{\mbox{T}}\:|0\rangle = |0\rangle.
\end{equation}

It turns out  that the unitary option has to be abandoned, as expected on the basis of general experience with other models. 
To see this, let us  assume that  $\hat{\mbox{T}}$ is unitary, hence linear.  Operators $\hat{b}^{\alpha}(\mathbf{p})$ in (47)  are expressed by $\hat{d}^{\alpha}(\mathbf{p})$
according to formula (31). Simple algebraic calculations shows that 
\[
	\hat{\mbox{T}}\: \hat{d}_{1}(\mathbf{p}) \hat{\mbox{T}}^{-1} = -  \eta_{\mbox{T}}\: i \: \hat{d}^{\dagger}_{2}(- \mathbf{p}), \;\;\; \hat{\mbox{T}}\: \hat{d}_{2}(\mathbf{p}) \hat{\mbox{T}}^{-1} =   \eta_{\mbox{T}}\: i \: \hat{d}^{\dagger}_{1}(- \mathbf{p}).
\]
Here the linearity of the operator $\hat{\mbox{T}}$ has been used in order to move this operator close to 
$\hat{d}_{\lambda}(\mathbf{p})$, to the position as on the l.h.s.'s of the formulas right above. We see that  condition (48) is not consistent with these formulas, e.g.,   $	\hat{\mbox{T}}\: \hat{d}_{1}(\mathbf{p}) \hat{\mbox{T}}^{-1} |0\rangle =0$ while $\hat{d}^{\dagger}_{2}(- \mathbf{p}) |0\rangle = |-\mathbf{p}\: 2\rangle \neq 0$. 

On the other hand, antiunitary $\hat{\mbox{T}}$ is antilinear, hence, when moving this operator close to $\hat{d}_{\lambda}(\mathbf{p})$, one has to complex conjugate  coefficients encountered on the way. In this case,  we obtain
\begin{equation}
	\hat{\mbox{T}}\: \hat{d}_{1}(\mathbf{p}) \hat{\mbox{T}}^{-1} =   \eta_{\mbox{T}}\: i \: \hat{d}_{2}(- \mathbf{p}), \;\;\; \hat{\mbox{T}}\: \hat{d}_{2}(\mathbf{p}) \hat{\mbox{T}}^{-1} =  - \eta_{\mbox{T}}\: i \: \hat{d}_{1}(- \mathbf{p}).
\end{equation}
These formulas are consistent with (48), and we may proceed with the definition of the operator $\hat{\mbox{T}}$ as in the case of space inversion $\hat{\mbox{P}}$. 

Now, equipped with the definition of the operator $\hat{\mbox{T}}$, we return to formulas (49)
which change their status: from a conjecture to easy to prove theorem. 
We can derive also the transformation law of the field operator  $\hat{\psi}$. It   resembles formula (46) for the classical field,
\begin{equation}  \hat{\mbox{T}} \hat{\psi}(\mathbf{x},t) \hat{\mbox{T}}^{-1} = \eta_{\mbox{T}} \gamma^0 \gamma_5 \hat{\psi}(\mathbf{x}, -t). 
\end{equation}	
The field operator is given by formula (35). When computing the l.h.s. in (50) one has to remember about the complex conjugation due to antiunitarity of $\hat{\mbox{T}}$. The matrix
$\gamma^0 \gamma_5$ is recovered  with the help of following formulas
\[ v_1 = i \gamma^0 \gamma_5 v_2^*,  \;\; v_2 = - i \gamma^0 \gamma_5 v_1^*,  \;\;  v_1^* = -i \gamma^0 \gamma_5 v_2,  \;\;  v_2^* =  i \gamma^0 \gamma_5 v_1.   \]

\section{Summary and remarks}

We have shown how to quantize the classical Majorana field starting from its expansion into the eigenfunctions of the axial momentum. 
 The model has the particle interpretation with a single spin 1/2 fermion. There is no anti-particle, as expected in the case of  Majorana field. The constructed quantum model essentially coincides with results of other approaches to quantization of the field, which is the desired outcome. Also the discrete symmetries: unitary $\hat{\mbox{P}}$ and antiunitary $\hat{\mbox{T}}$, have been implemented in the model  (the charge conjugation is trivial, $\hat{\mbox{C}}=I$). It is clear that the same eigenfunction expansion can be used also in quantum theory of the Dirac field.  

As for differences with other approaches to quantization of the Majorana field: (a) Our approach is self-contained -- we do not refer to the quantized Dirac field.  (b)   The mode expansion (5) is novel.  (c)  On a more technical level,  we have made the  specific  choice of the basis bispinors $e_{\alpha}(\mathbf{p})$, as described above formula (6). Due to it,  the coefficient functions $b^{\alpha}(\mathbf{p})$  in formula (5) have clear relativistic transformation laws (6), (7).   The concrete form of basis  bispinors $v_{\lambda}$ in formula (35) is a consequence of that choice. 

The presented quantization is not  based on canonical formalism. Instead, we have used as the guiding principle  the relativistic invariance. In particular, the operator 
$:\!\!\hat{P}_{0}\!\!:$ has been introduced as generator of time translations, without reference to a correspondence with a classical Hamiltonian.  Noteworthy is also the fact that there is no need to consider the Dirac sea when searching for the vacuum state.   

The fact that the expansion (5) has led to the expected quantum model confirms that the axial momentum is a useful quantum mechanical observable, in spite of its apparent peculiarities discussed in \cite{A2}, \cite{A3}.  

On a more general ground, we think that probably the most interesting aspect of our work is the appearance of the Majorana basis of operators, $\hat{b}^{\alpha}(\mathbf{p})$, in the context of quantized  relativistic fields. These operators were  replaced by the annihilation and creation operators $\hat{a}_{\lambda}(\mathbf{p}), \: \hat{a}^{\dagger}_{\lambda}(\mathbf{p})$ in Section 4 with the goal of obtaining the standard  Fock space and the particle interpretation. That construction provides  a particular complex representation of the Majorana basis of operators. 
A very interesting question arises about existence and features of other representations.

\section*{Appendix A. The  Lorentz boosts and the Wigner rotations}

Calculations in Sections 3 and 5 require  detailed knowledge of the Lorentz boosts and  of the Wigner rotations. For convenience of the reader,   we have collected  relevant formulas in this Appendix. Most of them are well-known, perhaps except formula $(A2)$ which is found in the monograph \cite{BLT}.

The proper orthochronous Lorentz matrices in a vicinity of the unit matrix $I_4$ can be written in the exponential form,  $L=\exp(\omega)$, where  $\omega$ is four by four  real matrix. Its elements $ \omega^{\mu\;\;}_{\;\;\nu}$ obey the condition   $\omega_{\mu\nu}= - \omega_{\nu\mu}$, where $\omega_{\mu\nu} = \eta_{\mu \kappa} \omega^{\kappa\;\;}_{\;\;\nu}$, $\eta_{\mu \kappa}$ are components of the Minkowski metric.  With this parametrization, $S(L) = \exp(\omega_{\mu\nu} [\gamma^{\mu}, \gamma^{\nu}]/8)$. In the Majorana representation, the matrices $\gamma^{\mu}$ are purely imaginary, $S(L)$ are real, and  $S(L)^T \gamma^0 = \gamma^0 S(L)^{-1}$. As the independent parameters on the Lorentz group we take $\omega^{23}, \omega^{31}, \omega^{12}, $ and $\omega^{0i}$ with $i=1,2,3.$

Let us stress that we rise or lower indices using the Minkowski metric, for example,
$\omega^{0\;\;}_{\;\;i} = - \omega^{0i}$. Moreover, the  $\delta_{ik}$ denotes the Kronecker symbol (not tensor), which always takes values 0 or +1. In this notation,  the trivial Lorentz transformation $L=I_4$ has matrix elements  denoted as  $(I_4)^{\mu\;\;}_{\;\;\nu}$, where  
$(I_4)^{0\;\;}_{\;\;0}=1,\; (I_4)^{i\;\;}_{\;\;k} = \delta_{ik}, \; (I_4)^{i\;\;}_{\;\;0} =0, \; (I_4)^{0\;\;}_{\;\;k}=0.  $
All this may sound trivial, but the reality is that in the calculations reported below it is very easy to make  a  sign mistake  related to the level of indices.

The matrix elements  $H^{\mu\;\;}_{\;\;\nu}(\mathbf{p})$ of the boost $H(\mathbf{p})$ have the following form:
\begin{equation*} H^{0\;}_{\;\;0}(\mathbf{p}) = \frac{E_p}{m}, \;  H^{0\;}_{\;\;i}(\mathbf{p}) = H^{i\;}_{\;\;0}(\mathbf{p})= \frac{p^i}{m}, \;  H^{i\;}_{\;\;j}(\mathbf{p})= \delta_{ij} + \frac{p^i p^j}{m(m+E_p)}. \eqno (A1) \end{equation*}
The matrix $H(\mathbf{p})$ is symmetric. It turns out that $(H(\mathbf{p}))^{-1} = H(-\mathbf{p})$.
In the case of this boost, 
 \begin{equation*} S(H(\mathbf{p}))= n_{\mu}(\mathbf{p}) \gamma^{\mu} \gamma^0, \eqno (A2) \end{equation*}
  where
\begin{equation*} n_0(\mathbf{p}) = \frac{m +E_p}{\sqrt{2m(m+E_p)}}, \;\; n_i(\mathbf{p}) = \frac{p_i}{\sqrt{2m(m+E_p)}}, \eqno (A3)\end{equation*}
see Exercise 2.4.9 in \cite{BLT}. Note that $p_i = - p^i$, $\mathbf{p} = (p^i)$. 

In Section 3 we use infinitesimal form of the Wigner rotations 
\[{\cal R}(L,\mathbf{p}) = H^{-1}(\mathbf{p}) \: L\: H(l^{-1}(\mathbf{p})).\]
In the case of rotations, the Lorentz matrix has the block diagonal form
\[L_R = \left(  \begin{tabular}{r|ccc} 1 &0 &0&0 \\ \hline  0&&& \\ 0&& {\large $R$}&\\ 0&&&  \end{tabular}  \right).\]
It turns out that \[{\cal R}(L_R,\mathbf{p}) = L_R.  \] 
In the linear approximation around $I_4$, which is sufficient for our purposes,
$ L_R \cong I_4 + \omega, $ where $\omega = ( \omega^{\mu\;\;}_{\;\;\nu})$ with $\omega^{0\;\;}_{\;\;i} =  \omega^{i\;\;}_{\;\;0} =0$, and 
\[ S(L_R) \cong I_4 + \sum_{i<k} \omega^{ik} [\gamma^i, \gamma^k] /4, \;\;\;  l^{-1}(\mathbf{p})^i \cong p^i + \omega^{ik} p^k.    \]
These formulas are used on the r.h.s. of formula (12) in the derivation of condition (13).   

In the case of infinitesimal boosts, $ L \cong I_4 + \delta L$,  where 
\[ \delta L = \left(  \begin{tabular}{r|ccc} 0 &$\omega^{0\;\;}_{\;\;1}$
&$\omega^{0\;\;}_{\;\;2} $&$\omega^{0\;\;}_{\;\;3} $ \\ \hline  $\omega^{1\;\;}_{\;\;0}$ &&& \\ $\omega^{2\;\;}_{\;\;0}$ && {\large 0}&\\ $\omega^{3\;\;}_{\;\;0} $&&&  \end{tabular}  \right).\]
Furthermore, $l^{-1}(\mathbf{p})^i \cong p^i - \omega^{i\;\;}_{\;\;0} E_p,   $  and
$ {\cal R}(L,\mathbf{p}) \cong I_4 + \delta{\cal R}, $ where \[\delta{\cal R}^{0\;\;}_{\;\;\mu} = 0,\;\;\; \delta{\cal R}^{\mu\;\;}_{\;\;0} = 0,\;\;\; \delta{\cal R}^{i\;\;}_{\;\;k} \cong \frac{1}{E_p +m} (\omega^{i0} p^k - \omega^{k0}p^i). \]  
The  first term on the r.h.s. of formula (15) comes from  
\[ S({\cal R}(L,\mathbf{p})) \cong I_4 + \frac{1}{8} \delta{\cal R}_{ik}\:[\gamma^i, \gamma^k],\] 
and the second term  from $\sqrt{E_{L^{-1}p }/E_p}  \;   b^{\alpha}(l^{-1}(\mathbf{p})). $

\section*{Appendix B. The commutator equations}

The formulas $(B2-B5)$ shown below facilitate calculations in Sections 3 and 4. 
 We seek a Hermitian operator $\hat{X}$ which obeys the condition 
\[  [\hat{X}, \hat{b}^{\alpha}(\mathbf{p})] = i \int\! d^3\mathbf{q}\; x_{\alpha\beta}(\mathbf{p}, \mathbf{q}) \:\hat{b}^{\beta}(\mathbf{q}),  \eqno (B1) \] 
where the real valued functions $x_{\alpha\beta}$ are antisymmetric in the following sense
	\[x_{\alpha\beta}(\mathbf{p}, \mathbf{q})  = - x_{\beta\alpha}(\mathbf{q}, \mathbf{p}). \]
 The condition $(B1)$ is satisfied by the Hermitian operator
\[\hat{X} = - \frac{i}{2}  \int\! d^3\mathbf{p} d^3\mathbf{q} \;   \hat{b}^{\alpha}(\mathbf{p}) x_{\alpha\beta}(\mathbf{p}, \mathbf{q}) \:\hat{b}^{\beta}(\mathbf{q}) + c_0 I, \eqno (B2) \]
where $c_0$ is an arbitrary real constant. This can be checked with the help of  the anticommutators (8) and the formula $[\hat{A}\hat{B},\hat{C}] = \hat{A}\: [\hat{B},\hat{C}]_+ -  [\hat{A},\hat{C}]_+\: \hat{B}$.

Let 
\[\hat{Y} = - \frac{i}{2}  \int\! d^3\mathbf{p} d^3\mathbf{q} \;  \hat{b}^{\alpha}(\mathbf{p}) y_{\alpha\beta}(\mathbf{p}, \mathbf{q})  \:\hat{b}^{\beta}(\mathbf{q}) + d_0 I, \]
where $d_0$ is an arbitrary real constant, and  the real valued functions $y_{\alpha\beta}(\mathbf{p}, \mathbf{q})$ are antisymmetric in the above sense. The commutator of the Hermitian operators  $\hat{X}$ and $\hat{Y}$ is given by the formula
\[ [\hat{X}, \hat{Y}] = - \frac{1}{2}  \int\! d^3\mathbf{p} d^3\mathbf{q} \;  \hat{b}^{\alpha}(\mathbf{p}) w_{\alpha\beta}(\mathbf{p}, \mathbf{q}) \:\hat{b}^{\beta}(\mathbf{q}),  \eqno (B3)  \]
where
\[ w_{\alpha\beta}(\mathbf{p}, \mathbf{q}) =  \int\! d^3\mathbf{s} \left( x^{}_{\alpha\eta}(\mathbf{p}, \mathbf{s}) y_{\eta\beta}(\mathbf{s}, \mathbf{q}) - y_{\alpha\eta}(\mathbf{p}, \mathbf{s}) x_{\eta\beta}(\mathbf{s}, \mathbf{q})      \right). \eqno (B4)  \]
Thus, the `matrix' $w$ for the commutator  is given by the commutator of the `matrices' $x$ and $y$. 

Commutators of the Poincar\'e generators discussed in Section 4 can be checked with the help of formula $(B5)$ below.  We consider operators 
$\hat{W}$  and $\hat{Z}$, 
\[ \hat{W}= \int \! d^3p d^3q \;\hat{d}^{\dagger}_i(\mathbf{p})  w_{ik}(\mathbf{p}, \mathbf{q})  \hat{d}_i(\mathbf{q}), \;\;
\hat{Z}= \int \! d^3p d^3q  \;\hat{d}^{\dagger}_i(\mathbf{p})  z_{ik}(\mathbf{p}, \mathbf{q})  \hat{d}_i(\mathbf{q}), \]
where $i,k =1,2$, and $ w_{ik},  z_{ik}$  are (generalized) functions of $\mathbf{p}, \mathbf{q}$.  Their commutator has the following form 
\[   [ \hat{W}, \hat{Z}] =    \int \! d^3p d^3q \;\hat{d}^{\dagger}_i(\mathbf{p}) r_{ik}(\mathbf{p}, \mathbf{q})  \hat{d}_i(\mathbf{q}), \eqno(B5)\]  
where
\[   r_{ik}(\mathbf{p}, \mathbf{q}) =    \int\! d^3\mathbf{s} \left( w^{}_{ij}(\mathbf{p}, \mathbf{s}) z_{jk}(\mathbf{s}, \mathbf{q}) - z_{ij}(\mathbf{p}, \mathbf{s}) w_{jk}(\mathbf{s}, \mathbf{q})      \right). \]

\end{document}